\documentclass{cccg24}
\usepackage{graphicx,amssymb,amsmath,hyperref,pgfplots}
\usepackage[normalem]{ulem}
\definecolor{periwinkle}{rgb}{0.8, 0.8, 1.0}
\definecolor{electricultramarine}{rgb}{0.25, 0.0, 1.0}\pgfplotsset{width=8cm,compat=1.9}

\title{Ruler Rolling}

\begin{document}
\thispagestyle{empty}
\maketitle

\begin{abstract}
At CCCG '21 O'Rourke proposed a variant of Hopcroft, Josephs and Whitesides' (1985) NP-complete problem {\sc Ruler Folding}.  For normal {\sc Ruler Folding}, we are asked to fold a carpenter's ruler whose segments have given integer lengths into an interval of at most a given length, alternating folds between 180 degrees clockwise and 180 degrees counter-clockwise.  For O'Rourke's variant, which he called {\sc Ruler Wrapping}, all folds must be 180 degrees in the same direction.  Gagie, Saeidi and Sapucaia (2023) noted that if the last straight section of the ruler must be longest, then {\sc Ruler Wrapping} is equivalent to partitioning a string of positive integers into substrings whose sums are increasing such that the last substring sums to at most a given amount.  They gave linear-time algorithms for the versions of {\sc Ruler Wrapping} both with and without this assumption.

In real life we cannot repeatedly fold a carpenter's ruler 180 degrees in the same direction.  In this paper we propose the more realistic problem of {\sc Ruler Rolling}, in which we repeatedly fold the segments 90 degrees in the same direction and thus fold the ruler into a rectangle instead of into an interval.  We should report all the Pareto-optimal rollings.  We note that if the last straight section of the ruler must be longer than the third to last --- analogously to Gagie et al.'s assumption --- then {\sc Ruler Rolling} is equivalent to partitioning a string of positive integers into substrings such that the sums of the even substrings are increasing, as are the sums of the odd substrings.

We give a simple dynamic-programming algorithm that reports all the Pareto-optimal rollings in quadratic time under this assumption.  Our algorithm still works even without the assumption, but then we are left with a quadratic number of two-dimensional feasible solutions, so finding the Pareto-optimal ones and increases our running time by a logarithmic factor. If we have a nice objective function, however, we still use quadratic time.
\end{abstract}

\section{Introduction}
\label{sec:introduction}

In 1985 Hopcroft, Joseph and Whitesides~\cite{HJW85} introduced the problem of {\sc Ruler Folding}: given the lengths of the segments in a carpenter's ruler each of whose hinges can be left straight or folded 180 degrees, with the directions of the folds alternating between clockwise and counter-clockwise, find the length of the shortest interval into which the ruler can be folded.  This is equivalent to considering the segments' lengths as a string of positive numbers and assigning each a sign $+$ or $-$ such that the difference between the minimum and maximum partial sums is minimized.  They showed that the decision version of this problem is NP-complete in the weak sense, via a reduction from {\sc Partition}, and gave a pseudo-polynomial time algorithm for it.  C\u{a}linescu and Dumitrescu~\cite{CD05} later gave a fully polynomial-time approximation scheme for it.

At the open-problem session of CCCG '21, O'Rourke~\cite{ORo21} proposed a variant of {\sc Ruler Folding} that he called {\sc Ruler Wrapping}, for which all the folded hinges must be folded 180 degrees in the same direction and we want to find the length of the shortest interval into which the ruler can be wrapped.  Gagie, Saeidi and Sapucaia~\cite{GSS23} noted that if the last straight section of the ruler must be the longest, then {\sc Ruler Wrapping} is equivalent to partitioning a string of positive integers into substrings whose sums are increasing such that the last substring sums to at most a given amount.  They gave simple, online and linear-time algorithms for the versions of {\sc Ruler Wrapping} both with and without this assumption, based on Knuth's algorithm~\cite{Knu73,Fre75} for {\sc Longest Increasing Subsequence}.

\begin{figure*}[t]
\begin{center}
\includegraphics[width=.6\textwidth]{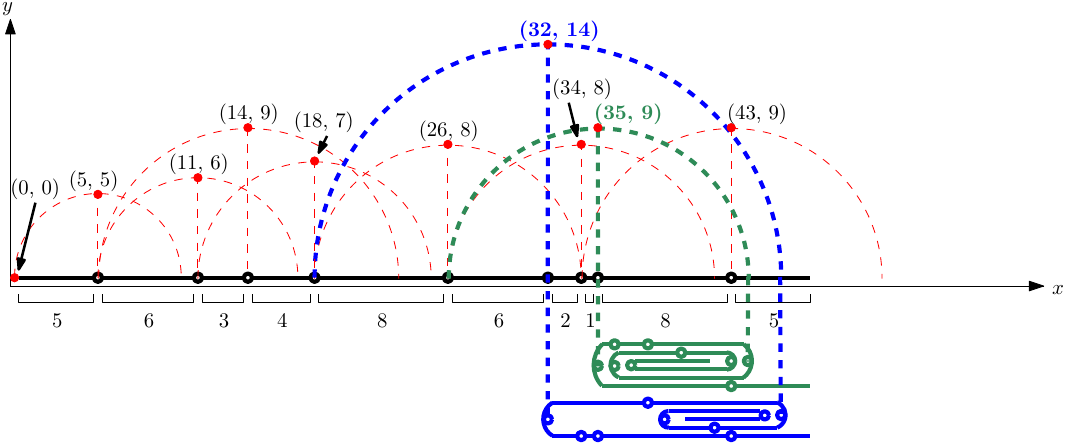}

\bigskip

\includegraphics[width=.6\textwidth]{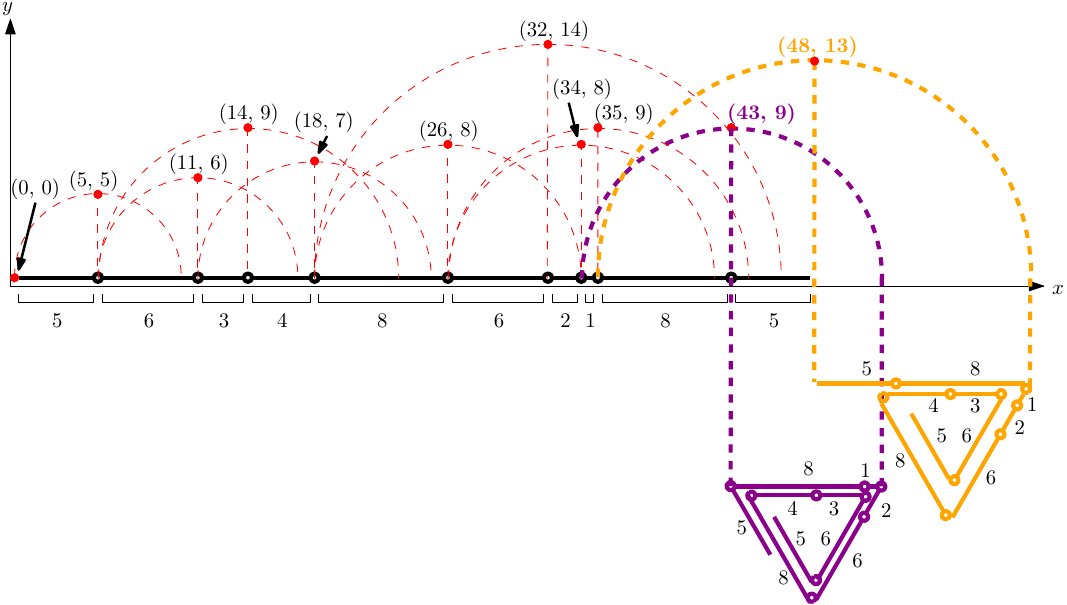}
\caption{Gagie et al.'s~\cite{GSS23} Figures 3 and 4.  When drawing their solutions as intervals {\bf (top)} became awkward, they eventually resorted to triangles {\bf (bottom)}.  The purple triangle is the optimal wrapping when the last straight section of the ruler can be as short as or shorter than the previous one, and the yellow triangle is the optimal wrapping when it must be longer.}
\label{fig:triangles}
\end{center}
\end{figure*}

As serious folding practitioners~\cite{Par11} know, however, the segments of a carpenter's ruler (and all other objects) have width as well as length, so repeatedly folding them 180 degrees in the same direction is problematic in real life.  Gagie et al.\ thus had difficulty even illustrating their solutions and eventually resorted to triangles, as shown in Figure~\ref{fig:triangles}.  In this paper we propose the more realistic problem of {\sc Ruler Rolling}, for which all the folded hinges must be folded 90 degrees in the same direction and we thus fold the ruler into a rectangle instead of into an interval.  Since a tall and thin rolling may be sometimes better and sometimes worse than a short and wide one, we want to list all the pairs $(h, w)$ such that the ruler can be rolled into a rectangle of height $h$ and width $w$ but cannot be rolled into a one of height $h'$ and width $w'$ with either $h' < h$ and $w' \leq w$ or $h' \leq h$ and $w' < w$.  In other words, we want the pairs corresponding to Pareto-optimal rollings.

We consider rolling rulers into rectangles instead of triangles because if the last straight section of the ruler must be longer than the third to last --- analogously to Gagie et al.'s assumption --- then {\sc Ruler Rolling} is equivalent to partitioning a string of positive integers into substrings such that the sums of the even substrings are increasing, and the sums of the odd substrings are increasing.  (We do not know of quite such a nice equivalence in the case of triangular rollings.)  We give a simple online dynamic-programming algorithm that reports all the Pareto-optimal rollings in quadratic time under this assumption.  Our algorithm still works without the assumption, but then it is not online and we are left with a quadratic number of feasible two-dimensional solutions, so finding the Pareto-optimal ones and discarding the others increases our running time by a logarithmic factor.  The running time drops back to quadratic, however, if we have a scalar objective function that can be computed in constant time and respects Pareto optimality in the sense that it assigns the same score to rollings with the same dimensions and assigns a better score to one rolling than to another if the former is shorter and not wider or thinner and not taller than the latter.  Intuitively, this is because the objective function projects all the solutions onto a line, and then we can find the minimum in time linear in the number of solutions and quadratic in the number of segments in the ruler.

In Section~\ref{sec:algorithm} we present our algorithm under the assumption that the last straight section of the ruler must be longer than the third to last.  More specifically, we assume the last segment is vertical and extends strictly below any other.  In Section~\ref{sec:correctness} we prove our algorithm's correctness.  In Section~\ref{sec:assumption} we discuss the challenges of dropping our simplifying assumption. possibly in favour of a nice objective function.  We leave as an open problem finding a quadratic-time algorithm for {\sc Ruler Rolling} with no assumptions at all.  In Section~\ref{sec:future} we discuss some other future work.

\section{Algorithm}
\label{sec:algorithm}

Suppose we are given the sequence $L = \ell_1, \ldots, \ell_n$ of the lengths of the $n$ segments in a carpenter's ruler and asked to return all pairs $(h, w)$ such that the ruler can be rolled with 90-degree folds in the same direction into a rectangle of height $h$ and width $w$ with the last segment vertical and extending strictly below every other, but cannot be rolled thus into a rectangle of height $h'$ and width $w'$ with either $h' < h$ and $w' \leq w$ or $h' \leq h$ and $w' < w$.  For an example, let us consider the ruler in Gagie et al.'s paper with $L = 5, 6, 3, 4, 8, 6, 2, 1, 8, 5$.

\begin{figure*}[t]
\begin{center}
\includegraphics[width=.8\textwidth]{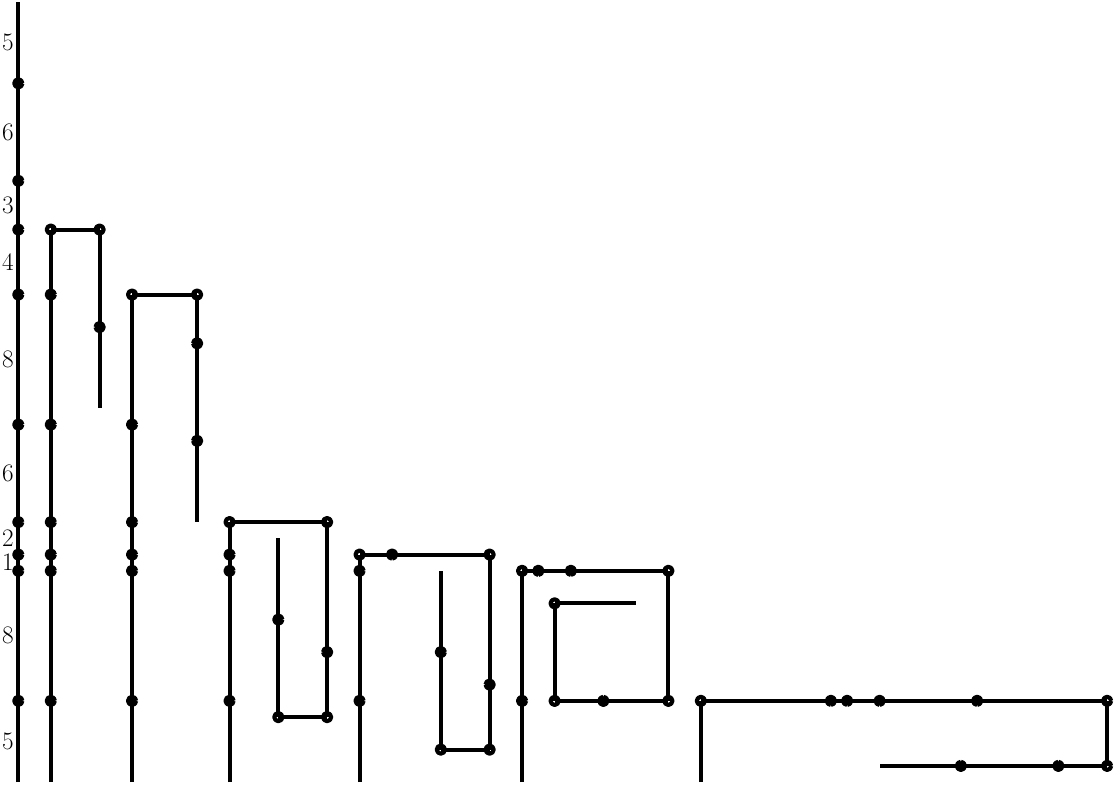}
\caption{The best rollings we can get for the ruler in Gagie et al.'s paper with $L = 5, 6, 3, 4, 8, 6, 2, 1, 8, 5$, with the last segment vertical and extending strictly below every other (so the first numbers in $L$ are the lengths of the innermost segments in the spirals).}
\label{fig:rollings}
\end{center}
\end{figure*}

To solve this problem, we build incrementally a table with $n + 1$ rows numbered from 0 to $n$, in which row $i$ stores $s_i = \sum_{j = 1}^i \ell_j$ and the pairs for the ruler with segment lengths $\ell_1, \ldots, \ell_i$, with the last of those segments vertical and extending strictly below every other.  Since we obtain the pairs for this ruler before looking at the following segments' lengths, our algorithm is online.  For our example, the last row of the table is row 10 and contains the pairs
\[(48, 0)\ \ (34, 3)\ \ (30, 4)\ \ (16, 6)\ \ (14, 8)\ \ (13, 9)\ \ (5, 25)\,,\]
corresponding to the rollings shown in Figure~\ref{fig:rollings}.  We note that the pair $(13, 9)$ could also correspond to a rolling with only 4 folds --- consider partitioning $L = 5, 6, 3, 4, 8, 6, 2, 1, 8, 5$ into $5, 6$ and $3$ and $4, 8$ and $6, 2, 1$ and $8, 5$ --- but the rolling with 5 folds shown in the figure is the one we find with our algorithm.

If we want the last segment to be horizontal instead of vertical then we can reverse the pairs, which rotates the rollings.  It follows that, if we can find in $O (n^2)$ time the dimensions of the Pareto-optimal rollings with the last segment vertical and extending strictly below every other, then we can find in $O (n^2)$ time the dimensions of the Pareto-optimal rollings with the last segment either horizontal or vertical and extending beyond every other in the relevant direction.

To solve the problem with the last segment vertical, we keep the pairs in each row sorted in decreasing order by first component and increasing order by second component.  The first six rows of the table for our example are below:
\[\begin{array}{r@{\hspace{.5ex}}c@{\hspace{0.5ex}}l|rrrr}
s_0 & = &  0 & (0, 0)\\
s_1 & = &  5 & (5, 0)\\
s_2 & = & 11 & (11, 0) & (6, 5)\\
s_3 & = & 14 & (14, 0) & (9, 5) & (3, 11)\\
s_4 & = & 18 & (18, 0) & (13, 5) & (7, 6) & (4, 14)\\
s_5 & = & 26 & (26, 0) & (12, 3) & (8, 7)
\end{array}\]

The first row of the table always contains $s_0 = 0$ and $(0, 0)$.  To compute row $i$ efficiently for $i > 0$, we first set $s_i = s_{i - 1} + \ell_i$.  For $0 \leq j < i$, we then
\begin{enumerate}
\item find the last pair $(h, w)$ in row $j$ such that $s_j + w < s_i$,
\label{item:find}
\item delete from the end of row $i$ any pairs whose second components are greater than $h$,
\label{item:delete}
\item append the pair $(s_i - s_j, h)$ to row $i$.
\label{item:append}
\end{enumerate}
In our example, we would compute row 6 as
\[\begin{array}{r@{\hspace{.5ex}}c@{\hspace{0.5ex}}l|rrrr}
s_6 & = & 32 & (32, 0) & (18, 3) & (14, 7) & (6, 12)
\end{array}\]
with $(27, 5)$ and $(21, 6)$ appended after $(32, 0)$ but then deleted again before we append $(18, 3)$.

Implemented na\"ively, our algorithm takes $O (n^3)$ time.  As $i$ increases, however, the position of the last pair $(h, w)$ in row $j$ such that $s_j + w < s_i$, is non-decreasing.  We can implement each row as a doubly-linked list and delete each pair whenever the next pair $(h, w)$ in the row has $s_j + w < s_i$, charging each pair's deletion to its insertion.  We also charge each pair's deletion in step~\ref{item:delete} to its insertion.  For each $j$ we append only one pair to row $i$ in step~\ref{item:append}, so steps~\ref{item:find} and~\ref{item:delete} take amortized constant time, and overall we use $O (n^2)$ time.

Readers may notice that, since $s_3 + 11 < s_i$ for $i \geq 5$ and the 3 in the pair $(3, 11)$ in row 3 is less than the 5 and 6 in the last pairs $(5, 0)$ and $(6, 5)$ in rows 1 and 2 --- and they must be the last pairs in their rows for this observation to work --- when computing rows 6 to 10 we can skip looking at rows 1 and 2, since the pairs we generate with them will always be deleted at least when we append the pair we generate with row 3.  We do not see how to use this observation to improve our worst-case bound, however.

The prefix of our table shown above is shown below on the left but without the pairs that are deleted or ignored after we have computed row 5.  When we compute row 6, we delete the pair $(26, 0)$ from row 5, as shown below.
\[\begin{array}{r@{\hspace{.5ex}}c@{\hspace{0.5ex}}l|rrrr}
s_0 & = &  0 & (0, 0)\\
s_1 & = &  5 & \\
s_2 & = & 11 & \\
s_3 & = & 14 & (3, 11)\\
s_4 & = & 18 & (7, 6) & (4, 14)\\
s_5 & = & 26 & (26, 0) & (12, 3) & (8, 7) \\[3ex]
\hline &&&&&&\\[0ex]
s_0 & = &  0 & (0, 0)\\
s_1 & = &  5 & \\
s_2 & = & 11 & \\
s_3 & = & 14 & (3, 11)\\
s_4 & = & 18 & (7, 6) & (4, 14)\\
s_5 & = & 26 & \mbox{\sout{$(26, 0)$}} & (12, 3) & (8, 7)\\
s_6 & = & 32 & (32, 0) & (18, 3) & (14, 7) & (6, 12)
\end{array}\]
Continuing like this, the pairs left before we compute row 10 are shown in Table~\ref{tab:last}, with the ones we delete then crossed out.

\begin{table*}[t]
\begin{center}
\[\begin{array}{r@{\hspace{.5ex}}c@{\hspace{0.5ex}}l|rrrrrrrr}
s_0 & = &  0 & (0, 0)\\
s_1 & = &  5 & \\
s_2 & = & 11 & \\
s_3 & = & 14 & (3, 11)\\
s_4 & = & 18 & (4, 14)\\
s_5 & = & 26 & \mbox{\sout{$(8, 7)$}}\\
s_6 & = & 32 & \mbox{\sout{$(14, 7)$}} & (6, 12)\\
s_7 & = & 34 & (8, 8) & (2, 32)\\
s_8 & = & 35 & \mbox{\sout{$(17, 4)$}} & (9, 8) & (3, 32) & (1, 34)\\
s_9 & = & 43 & \mbox{\sout{$(43, 0)$}} & \mbox{\sout{$(29, 3)$}} & (25, 4) & (9, 8) & (8, 17)\\
s_{10} & = & 48 & (48, 0) & (34, 3) & (30, 4) & \mbox{\sout{$(22, 8)$}} & (16, 6) & (14, 8) & (13, 9) & (5, 25)
\end{array}\]
\caption{The pairs left before we compute row 10, with the ones we delete then crossed out.}
\label{tab:last}
\end{center}
\end{table*}

We note that, even when we delete pairs from the beginnings of lists, they may be useful if we later want to find the actual Pareto-optimal rollings themselves.  To see why, consider Table~\ref{tab:full}, which shows all the pairs we generate for our example.  There are boxes around the pairs --- some of which are deleted or crossed out in Table~\ref{tab:last} --- that are partial solutions leading to the pair $(13, 9)$ in row 10 that corresponds to the Pareto-optimal sixth rolling from the left in Figure~\ref{fig:rollings}.  As usual with dynamic programming, we can find the actual rollings efficiently if we keep a pointer from each pair to the one from which we generated it.

\begin{table*}[t]
\begin{center}
\[\begin{array}{r@{\hspace{.5ex}}c@{\hspace{0.5ex}}l|rrrrrrrr}
s_0 & = &  0 & \fbox{(0, 0)}\\
s_1 & = &  5 & \fbox{(5, 0)}\\
s_2 & = & 11 & (11, 0) & \fbox{(6, 5)}\\
s_3 & = & 14 & (14, 0) & (9, 5) & (3, 11)\\
s_4 & = & 18 & (18, 0) & (13, 5) & \fbox{(7, 6)} & (4, 14)\\
s_5 & = & 26 & (26, 0) & (21, 5) & (15, 6) & (12, 3) & \fbox{(8, 7)}\\
s_6 & = & 32 & (32, 0) & (18, 3) & (14, 7) & (6, 12)\\
s_7 & = & 34 & (34, 0) & (20, 3) & (16, 4) & (8, 8) & (2, 32)\\
s_8 & = & 35 & (35, 0) & (21, 3) & (17, 4) & \fbox{(9, 8)} & (3, 32) & (1, 34)\\
s_9 & = & 43 & (43, 0) & (29, 3) & (25, 4) & (17, 8) & (11, 14) & (9, 8) & (8, 17)\\
s_{10} & = & 48 & (48, 0) & (34, 3) & (30, 4) & (22, 8) & (16, 6) & (14, 8) & \fbox{(13, 9)} & (5, 25)
\end{array}\]
\caption{All the pairs we compute, with boxes around the ones --- some of which are deleted or crossed out in Table~\ref{tab:last} --- that are partial solutions leading to the pair $(13, 9)$ in row 10 that corresponds to the Pareto-optimal sixth rolling from the left in Figure~\ref{fig:rollings}.}
\label{tab:full}
\end{center}
\end{table*}

\section{Proof of correctness}
\label{sec:correctness}

We prove the correctness of our algorithm by induction on the row number.  Row 0 is always the same and stores $s_0 = 0$ and $(0, 0)$.  Assume that we have computed rows 0 to $i - 1 \geq 0$ correctly and now we want to compute row $i$.  Any Pareto-optimal rolling $W$ of the ruler with segments lengths $\ell_1, \ldots, \ell_i$ either has a last folded hinge at distance $s_j$ from the beginning of the ruler, for some positive $j < i$, or has no folded hinges, in which case we generate its pair from $(0, 0)$.  Without loss of generality, assume $W$ has a folded hinge.

The side of $W$ that ends with the $i$th segment has length $s_i - s_j$ so --- since we require this $i$th and last segment to be vertical and extend below every other --- the height of $W$ is also $s_i - s_j$.  If we cut off that whole side and rotate $W$ 90 degrees counter-clockwise, then we obtain a rolling for the ruler with segment lengths $\ell_1, \ldots, \ell_j$ in which the $j$th and last segment is vertical and extends below every other.  For example, if we cut off the left size of the sixth rolling in Figure~\ref{fig:rollings}, which has pair $(13, 9)$, and rotate that rolling 90 degrees counter-clockwise then, as shown at the top and right of the figure, we obtain a rolling with pair $(9, 8)$ for the ruler with segment lengths $5, 6, 3, 4, 8, 6, 2, 1$.  Without loss of generality we can assume this new rolling is Pareto-optimal --- if it is not, there must be another that is and generates $W$, or $W$ itself could not be Pareto-optimal --- so the pair for it is in row $j$.  If this pair is $(h, w)$, then the height of $W$ is $h$, and $W$'s pair is the one $(s_i - s_j, h)$ we would generate for $\ell_1, \ldots, \ell_i$ from $(h, w)$.

\begin{figure}[t]
\begin{center}
\includegraphics[width=0.3\textwidth]{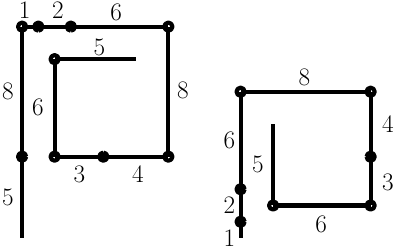}
\caption{If we cut off the left size of the sixth rolling in Figure~\ref{fig:rollings}, which has pair $(13, 9)$, and rotate that rolling 90 degrees counter-clockwise, then we obtain a rolling with pair $(9, 8)$ for the ruler with segment lengths $5, 6, 3, 4, 8, 6, 2, 1$.}
\label{fig:rotation}
\end{center}
\end{figure}

The last thing left for us to show is that we do in fact consider $(h, w)$ in row $j$ when building row $i$.  There are three possible reasons we might not:
\begin{enumerate}
\item $s_j + w \geq s_i$ and so we have not reached $(h, w)$ yet in row $j$;
\label{item:blocked}
\item $(h, w)$ is followed by another pair $(h', w')$ in row $j$ with $s_j + w' < s_i$ and so we have already deleted $(h, w)$;
\label{item:deleted}
\item $(h, w)$ is the last pair in row $j$ and some pair $(h'', w'')$ in a later row $j' < i$ has $h'' \leq h$ and $s_{j'} + w'' < s_i$, so we ignore $(h, w)$.
\label{item:ignored}
\end{enumerate}
If reason~\ref{item:blocked} holds then, even if we considered $(h, w)$, the part of $W$ between the last folded hinge and the $i$th segment would be shorter than the opposite side of $W$, so the $i$th segment would not extend below every other, as required.  If reason~\ref{item:deleted} holds then, since we keep the pairs in decreasing order by first component, $h' < h$ and so from $(h', w')$ we generate a rolling with the same height as $W$ and smaller width, contradicting $W$'s Pareto-optimality.  Similarly, if reason~\ref{item:ignored} holds, then we generate a rolling with height $s_i - s_{j'} < s_i - s_j$ and width $h'' \leq h$, again contradicting $W$'s Pareto-optimality.  Summing up, we have the following theorem:

\begin{theorem}
Given the lengths of the $n$ segments in a carpenter's ruler, in $O (n^2)$ time we can return all Pareto-optimal pairs $(h, w)$ such that the ruler can be rolled into a rectangle of height $h$ and width $w$ with the last segment extending strictly beyond every other in the relevant direction.
\end{theorem}

\section{Assumption}
\label{sec:assumption}

\begin{figure}[t]
\begin{center}
\includegraphics[width=0.47\textwidth]{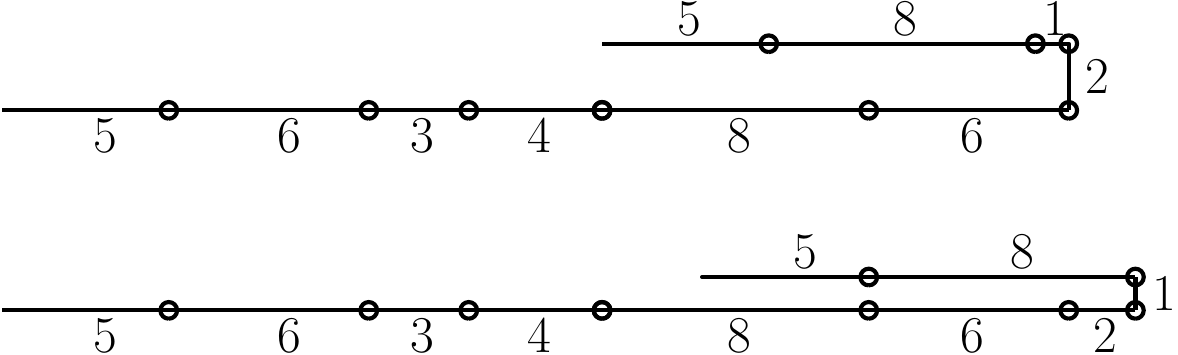}
\caption{The rollings corresponding to the pairs $(2, 32)$ and $(1, 34)$, left in rows 7 and 8, which do not have the last segment extending beyond every other.}
\label{fig:extras}
\end{center}
\end{figure}

The pairs we insert into other rows and never delete but do not consider when computing row $n$, correspond to rollings in which the last segment does not extend strictly below every other.  In our example, these are $(2, 32)$ in row 7, $(3, 32)$ and $(1, 34)$ in row 8, and $(9, 8)$ and $(8, 17)$ in row 9.  The pairs $(2, 32)$ and $(1, 34)$ are particularly interesting since, not only are they Pareto-optimal without our assumption, but $(2, 32)$ dominates $(3, 32)$ in row 10 and both $(2, 32)$ and $(1, 34)$ dominate $(34, 3)$ in row 10 when reversed.  The corresponding rollings are shown in Figure~\ref{fig:extras} (rotated 90 degrees as an example of how this can save space!).

Since our algorithm runs in $O (n^2)$ time and generates $O (n^2)$ pairs in total, in $O (n^2 \log n)$ time we can easily find the ones that are Pareto-optimal without our assumption --- but then our algorithm is not online and not quadratic.  On the other hand, if we have a scalar objective function that can be computed in constant time and assigns the same score to rollings with the same dimensions and assigns a better score to one rolling than to another if the former is shorter and not wider or thinner and not taller than the latter, than
\begin{itemize}
\item we are safe to delete the pairs that we delete, since they are dominated by other pairs;
\item we can evaluate the objective function on all $O (n^2)$ pairs that remain in our linked lists and find and report the ones with the best score in $O (n^2)$ time.
\end{itemize}
In other words, with such an objective function our running time drops back to quadratic.  We leave as an open problem finding a quadratic-time algorithm for {\sc Ruler Rolling} with no assumptions at all.

\begin{theorem}
Given the lengths of the $n$ segments in a carpenter's ruler, in $O (n^2 \log n)$ time we can return all Pareto-optimal pairs $(h, w)$ such that the ruler can be rolled into a rectangle of height $h$ and width $w$.  If we have a scalar objective function that can be computed in constant time and assigns the same score to rollings with the same dimensions and assigns a better score to one rolling than to another if the former is shorter and not wider or thinner and not taller than the latter, than we use $O (n^2)$ time.
\end{theorem}

\section{Future work}
\label{sec:future}

Apart from finding an algorithm for {\sc Ruler Rolling} that runs in quadratic time without assumptions, we think the most interesting open problem regarding {\sc Ruler Rolling} is determining whether quadratic time is necessary in the worst case.  We conjecture that it is necessary, at least for online algorithms, although we seem able to use much less time in practice when we skip looking at rows unnecessarily, as described in Section~\ref{sec:algorithm}.

To test our algorithm's running time in practice --- with the assumption that the last straight section is longer than the third to last --- we implemented two versions of it (available at \url{https://github.com/kevinlyu1006/Ruler-Wrapping-Problem-Implementation}), one unoptimized and the other with skipping.  Skipping clearly provided a dramatic speedup, as shown in Figure~\ref{fig:runtime_comparison} in the appendix.  During testing, we noticed that as the upper bound on the segment length increased, the run-time of the version with skipping decreased, as shown in Figure~\ref{fig:length_speedup} in the appendix, although there was no significant speedup for the unoptimized version

\begin{figure}[t]
\begin{center}
\includegraphics[width=.3\textwidth]{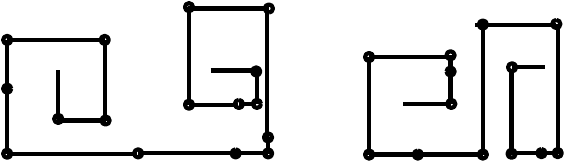}
\caption{Rollings in which we change from increasing to decreasing sums {\bf (left)} and change the folding direction {\bf (right)}.}
\label{fig:scrollings}
\end{center}
\end{figure}

In the full version of this paper we will also discuss rolling rulers into triangles, and the versions of {\sc Ruler Rolling} in which we can switch from increasing to decreasing sums or change the folding direction from clockwise to counter-clockwise and back, as illustrated in Figure~\ref{fig:scrollings}.  The former version reduces fairly easily to {\sc Ruler Rolling} but Hopcroft et al.'s reduction from {\sc Partition} to {\sc Ruler Folding} can fairly easily be modified to reduce to the latter version, so it too is NP-complete.

\newpage
\section*{Appendix}

\begin{figure}[h]
\begin{center}
\resizebox{.4\textwidth}{!}
{\begin{tikzpicture}
\begin{axis}[
    xlabel={$\log_{10} n$},
    ylabel={Run-time (ms)},
    xmin=0.5, xmax=5.5,
    ymin=0, ymax=1500000,
    ytick={0,3e5,6e5,9e5,12e5,15e5},
    scaled y ticks=false,
    y tick label style={font=\footnotesize},
    yticklabel style={/pgf/number format/sci},
    legend pos=north west,
    ymajorgrids=true,
    grid style=dashed
]

\addplot[
    only marks,
    mark=*,
    color=red,
    ]
    coordinates {
    (1,0.2)(1.5,0.4)(2,1.4)(2.5,13.5)(3,132.3)(3.5,1193.2)(4,11812.2)(4.5,120514.7)(5,1466289.2)
    };
\addlegendentry{unoptimized}

\addplot[
    only marks,
    mark=square*,
    color=blue,
    ]
    coordinates {
    (1,0)(1.5,0.2)(2,0.6)(2.5,3.5)(3,16.5)(3.5,93.9)(4,564.9)(4.5,3385)(5,23415.1)
    };
\addlegendentry{with skipping}

\end{axis}
\end{tikzpicture}}
\caption{The run-times of our unoptimized implementation {\bf (red circles)} and of our implementation with skipping {\bf (blue squares)}.  Each point indicates the average over 10 trials with $n$ pseudo-randomly chosen integers from 1 to 100 in each trial.  (We note that all but the leftmost two red circles are almost or totally hidden behind the corresponding blue squares.)}
\label{fig:runtime_comparison}

\bigskip

\resizebox{.4\textwidth}{!}
    {\begin{tikzpicture}
        \begin{axis}[
            xlabel={$\log_{10} l$},
            ylabel={Run-time (s)},
            xmin=0, xmax=3.5,
            ymin=0, ymax=100000,
            xtick={0,1,2,3},
            ytick={0,2e4,4e4,6e4,8e4,1e5},
            scaled y ticks=false,
            yticklabel style={/pgf/number format/sci},
            xticklabel style={font=\small},
            yticklabel style={font=\small},
            grid=both,
            grid style={line width=.1pt, draw=gray!10},
            major grid style={line width=.2pt,draw=gray!50},
            minor tick num=1,
            legend style={font=\small},
            scatter/classes={
                a={mark=square*,blue,mark size=2}
            },
        ]
        \addplot[scatter,only marks,scatter src=explicit symbolic]
            coordinates {
                (0.477, 92817.9) [a]
                (1, 54382.2) [a]
                (1.505, 31945.8) [a]
                (2, 23415.1) [a]
                (2.50, 19783.4) [a]
                (3, 18661.3) [a]
            };
        \end{axis}
    \end{tikzpicture}}
\caption{The average run-time in practice decreases as the upper bound on the segment length increases.  Each point indicates the average over 10 trials with $n = 100000$ pseudo-randomly chosen integers from 1 to the segment-length upper bound ($l$) in each trial.}
\label{fig:length_speedup}
\end{center}
\end{figure}

\end{document}